\documentclass[epj]{svjour}

\usepackage{comment}
\usepackage{graphicx}
\usepackage{dcolumn}
\usepackage{bm}
\usepackage{enumitem}
\usepackage{threeparttable}
\usepackage{rotating}

\usepackage{color}
\usepackage{multirow}

\newcommand{\fig}[1]{Fig.~\ref{#1}}
\newcommand{\tab}[1]{Table~\ref{#1}}
\newcommand{\Sec}[1]{Sec.~\ref{#1}}
\newcommand{\eq}[1]{Eq.~(\ref{#1})}
\newcommand{\Rref}[1]{Ref.~\cite{#1}}

\newcommand{\g}{$\gamma$}

\newcommand{\HeHe}{$^3$He($\alpha$,$\gamma$)$^7$Be }

\begin{document}

\title{Gas-cell development for nuclear astrophysics motivated studies on noble gas targets and the $^3$He($\alpha$,$\gamma$)$^7$Be reaction}%

\author{\'A.~T\'oth\inst{1,2} \and Z.~Elekes\inst{1,3} \and Zs.~F\"ul\"op\inst{1} \and Gy.~Gy\"urky\inst{1} \and Z.~Hal\'asz\inst{1} \and M.\,M.~Juh\'asz\inst{1}\thanks{Present address: Instituto de Física Corpuscular (CSIC), Valencia, Spain}
 \and G.\,G.~Kiss\inst{1} \and S.\,R.~Kov\'acs\inst{1,2} \and Zs.~M\'atyus\inst{1,2} \and T.\,N.~Szegedi\inst{1}\thanks{Present address: Istituto Nazionale di Fisica Nucleare - Laboratori Nazionali del Sud (INFN-LNS), Catania, Italy}
 \and T.~Sz\"ucs\inst{1}\thanks{e-mail: tszucs@atomki.hu (corresponding author)}}

\institute{
HUN-REN Institute for Nuclear Research (Atomki), PO Box 51, 4001 Debrecen, Hungary \and
Doctoral School of Physics, Faculty of Science and Technology, University of Debrecen, Egyetem tér 1., 4032 Debrecen, Hungary \and
Institute of Physics, Faculty of Science and Technology, University of Debrecen,  Egyetem tér 1., 4032 Debrecen, Hungary %
 }

\date{Received: date / Revised version: date   / version date: \today}

\date{\today}

\abstract{
In many astrophysical scenarios, alpha induced reactions on noble gas nuclei play a crucial role. Studying these reactions in the laboratory requires the noble gas atoms to be confined in a sufficient amount to allow the reactions.
At Atomki thin-windowed gas-cell targets were developed and improved for studying alpha induced reactions on noble gases.
Several stages of the gas-cell design used for activation experiments and lately a version to be used for particle scattering experiments will be presented. \\
A new experimental study of the  \HeHe reaction with one of the activation gas-cell targets was performed. This reaction plays an important role both in the solar pp-chains and in big bang nucleosynthesis. The reaction cross section was measured in the past in several works, however, there are still energy regions lacking experimental data, rendering the extrapolations towards the astrophysically relevant energies uncertain. New experimental total cross section of the \HeHe reaction was thus determined here in the energy range of \mbox{$E_\mathrm{c.m.} = 2600-3000$~keV} in about 50 keV energy steps. These results confirm the overall trend, and also the absolute scale set by the only one previous measurement in this energy range. \\
In addition, two pilot experiments with the scattering cell were performed aiming to study the $^4$He($\alpha$,$\alpha$)$^4$He and $^{124}$Xe($\alpha$,$\alpha$)$^{124}$Xe reactions at \mbox{$E_\alpha = 18$~MeV}. These studies benchmark the performance of the cell and detection system both for light and heavy noble gas targets.
}
\authorrunning{\'A.~T\'oth {\it et al.}}
\titlerunning{Gas-cell development for nuclear astrophysics motivated studies on noble gas targets}
\maketitle

\section{\label{sec:intro}Introduction}

Noble gas targets are highly important in astrophysics motivated nuclear experiments.
In several astrophysical scenarios, both of the reaction partners are noble gas nuclei. 

For example, the \HeHe reaction is important both in the big bang nucleosynthesis (BBN) \cite{Coc17-IJMPE} and in the solar pp-chains \cite{Acharya25-RMP}. In the former scenario, the reaction is the main producer of $^7$Li. In BBN model calculations the abundance of only this isotope does not match the observations, often addressed as the primordial lithium problem \cite{Fields11-ARNPS}. In case of the Sun, the reaction is the branching towards the neutrino producing chains, thus directly influences the neutrino flux, i. e. the modeled $^7$Be neutrino flux uncertainty is dominated by the rate uncertainty of \HeHe reaction. A more precise determination of the cross section may lead to a reduced rate uncertainty, thus to a more precise modeled flux.

In case of the heavy element synthesis, large reaction networks have to be considered, for which the rates are usually derived from theoretical cross section calculations, mainly using the Hauser-Feshbach statistical model \cite{Hauser52-PR}. Experiments on proton and alpha induced reactions are important to pin down the model parameters. In many cases, reactions on noble gas isotopes have key importance in these parameter determinations. As an example, the $^{124}$Xe($\alpha$,$\gamma$)$^{128}$Ba reaction is a key reaction in the p-process reaction network \cite{Rauscher06-PRC,Rapp06-AJ}, while $^{86}$Kr($\alpha$,n)$^{89}$Sr is important in the weak r-process \cite{Bliss20-PRC,Psaltis22-AJ}.

If proton induced reactions are considered, the limitations of a noble gas target may be mitigated with inverse kinematic studies, where the beam is the heavier noble gas reactant impinging on hydrogen containing target \cite{Glorius19-PRL}. However, in case of alpha induced reactions, both of the reaction partners are noble gases, thus exchanging them still does not solve the difficulties to obtain noble gas target.
In a typical nuclear astrophysics experiment the target needs to have an areal density at least in the order of the $10^{16}$~atoms/cm$^2$, but in some cases even $10^{19}$ atoms/cm$^2$ is required. This thickness is easily achieved with solid state targets, however, noble gases hardly form solid compounds. At the same time the targets have to withstand an ion beam bombardment of a few Coulombs with a rate of a few hundreds of nA up to a few tens of $\mu$A \mbox{($10^{12} - 10^{14}$~particle/s)} without major deterioration to obtain measurable reaction rate. In addition, in many cases isotopically enriched materials have to be used, which could be quite expensive or simply not available in higher quantities, thus posing a strong constraint on the amount of total material to be used.
One way to create solid targets out of noble gases is implantation, when accelerated noble gas atoms impinge on a solid host material, and are trapped in the crystalline structure. Both the stability and the achievable thickness of such targets depend heavily on the used host material. Even though the thickness can reach the minimum required $10^{16}$ atoms/cm$^2$, in many cases -- due to the beam induced heat -- the noble gas atoms may escape, causing deterioration of the target \cite{Lhuillier11-JNM}.
Alternatively, thin targets can be made by sputtering material in noble-gas atmosphere, forming films with closed cell pores containing the gas on the host surface. However, the target thickness in these case is usually not sufficient for a typical nuclear astrophysics experiment, and opening of the closed cells due to the deposited beam power can cause uncontrolled degradation \cite{Ibrahim24-V}.

With the use of gas targets in most cases the above limitations can be mitigated, however, the confinement of the gas in the beam-line vacuum, while letting the beam interact with it, poses its own challenges. 
 There are in general three ways to achieve this. In the first method the outflow of the gas from the interaction volume is restricted by small apertures while the gas is continuously resupplied in order to compensate the loss (windowless extended gas target)  \cite{Ferraro18-EPJA,Paneru24-PRC}. The second option is dynamic confinement where the continuously supplied gas has no time to escape from a restricted volume owing to its high speed (gas jet target) \cite{Schmidt18-NIMA,Yadav26-NIMA}, and the third is static confinement with windows (thin windowed gas-cell target) \cite{Bordeanu12-NIMA,Toth23-PRC}.
In the first two cases, staged pumping is necessary to reach the vacuum levels required for the accelerated ion beam, thus a complex system of pumps is inevitable. In addition, recirculation is usually mandatory, which may include a gas purifier for removing any in-leaked contaminants and a buffer volume to maintain a steady gas flow, for the gas jet even a compressor stage is likely necessary.
Considering few hundreds of millibar pressure in the buffer and in the purifier, a minimum of about a standard liter of gas is necessary for a given experiment with windowless extended gas target \cite{Cavanna14-EPJA}, and even more for a gas jet target \cite{Yadav23-EPJWC}.
In contrast, in case of a gas-cell target with a typical few tens of a cm$^3$ overall volume and few hundreds of a millibar pressure, only a few hundredths of a standard liter gas is necessary. In addition, no complex pumping stages are required, since there is no continuous gas flow, which has to be dealt with.

The purpose of this paper is to present the stages of the development of thin-window gas-cell target systems and their application to the measurement of the \HeHe reaction cross section. In addition, gas-cell target development towards the measurement of $^3$He+$^4$He scattering is also shown. The paper is organized as follows: in \Sec{sec:data_analysis} general introduction about data analysis specialties with gas-cell targets is given. In \Sec{sec:cells} the design and main parameters of the gas-cell targets are detailed. This is followed by description of experiments towards particle scatting studies with gas-cell in \Sec{sec:scatt}, while the new \HeHe experiment and its results are presented in \Sec{sec:hehe}. Finally a summary and outlook is given in \Sec{sec:sum}.

\section{\label{sec:data_analysis}Cross section determination with gas-cells}

\subsection{\label{sec:gen}General considerations}

In practical terms, there are only small differences in deriving cross section data from measurements with gas-cell and solid state targets using the activation method \cite{Gyurky19-EPJA}.
In case of a gas-cell, the number of beam particles can be determined from charge integration, identically to a solid target experiment, if the complete gas-cell including the gas handling is part of an electrically isolated Faraday-cup. In this way, the possible charge exchange processes happening inside the gas or foils do not affect the current reading as opposed to other types of gas targets.
The number of target atoms can be derived from the initial pressure, temperature and gas-cell length, using the ideal gas law. Since the gas-cell is sealed, the areal number density of the active target atoms do not change, even if the temperature and/or pressure changes due to the heating effect of the beam. Gas can desorb from the inner gas-cell surfaces, however, that is either air or water vapor absorbed earlier, when the gas-cell was exposed to the ambient atmosphere. The extra gas may change the effective beam energy and collection efficiency of the reaction products, but does not change the number of active target atoms.
In case of a solid target activation experiment, the reaction products are naturally embedded into the target or backing, thus its activity can be measured. However, in the case of an activation gas target, these are created along the beam path in the gas volume, and have to be collected for the activity counting. The motion of the reaction products are governed by the reaction kinematics, and usually have a forward momentum which is made use for capturing them. Usually a sheet of metal, the so-called catcher, is placed at the back end of the gas volume into which the reaction products are implanted. Since this catcher contains the radioactive reaction products, they can be handled after the irradiation as the solid activation targets.

In case of particle scattering in a gas-cell (see \Sec{sec:scatt}), some of the above considerations are not relevant, because the scattering cross section is usually measured relative to a given angle (monitor detector) where the Rutherford cross section dominates \cite{Mohr13-ADNDT}. Both the target thickness and target current factor out. Even a slight target loss can be allowed during the measurement, not affecting the data analysis. It is important, however, that the scattered particles have to reach the detectors, and the scattering angle has to be well determined.

\subsection{\label{sec:beam_width}Effect of the energetic beam width}

The main difference between a solid target experiment and a gas-cell which shall be considered is the energy loss in the entrance foil and the beam energy spread. The latter includes the initial energetic beam width in addition to the energetic widening caused by beam energy straggling in the entrance foil.

The standard relation (considering the stopping power to be constant within the target) \cite{Iliadis-book} connecting the experimental yield ($Y$) with the reaction cross section ($\sigma$) is the following:
\begin{equation}
\label{eq:XS}
Y(E_0) = n  \sigma(E_\mathrm{eff.}),
\end{equation}
where $E_0$ is the beam energy at the target surface, $n$ is the areal number density of the target atoms, while $E_\mathrm{eff.}$ is a specific energy value attributed to the deduced cross section value.
If $\sigma$ is strictly constant, this energy can be anything from $E_0-\Delta E$ to $E_0$, however, usually \mbox{$E_\mathrm{eff.} = E_0-\Delta E / 2$} as the energy at the middle of the target.
\eq{eq:XS} is still valid if the cross sections is slightly varying\footnote{i. e. the $\sigma$ value does not change significantly within the energy range of interest. This significance is not well defined in the literature, however, \cite{Lemut08-EPJA} for example suggest a maximum 1\% change.} within the energy range covered by the target. In this case, $E_\mathrm{eff.} = E_0-\Delta E / 2$ is still a good approximation, and we will use this definition in the following despite the possible other definitions shown e.~g. in \cite{Lemut08-EPJA}.

If substantial beam energy spread is present, the following more general expression shall be used instead \cite{Rolfs-book}:
\begin{equation}
\label{eq:yield}
Y(E_0) = n \int_{E_0-\Delta E}^{E_0}  \int_{0}^{\infty} \sigma(E') g(E',E_0) dE' dE,
\end{equation}
where $g$ is the normalized beam energy distribution. This equation still assumes that the beam energy distribution does not change significantly along the target.
In practical cases, it is sufficient to approximate $g$ with a Gaussian function centered at the beam energy $E_0$, when that enters the active volume, and having a given standard deviation, which in the following will be called as the energetic beam width.

If $\sigma$ is slightly varying, the above equations reduces to the following relation
\begin{equation}
\label{eq:XS_int}
Y(E_0) = n  \sigma(E_\mathrm{eff.}) \int_{0}^{\infty} \mathrm{\Pi}_{E_0-\Delta E,E_0}(E) g(E_0-E) dE,
\end{equation}
where $\mathrm{\Pi}_{E_0-\Delta E,E_0}$ represents a boxcar function with a width of $\Delta E$.
The integral is apparently the convolution of the beam distribution with the target thickness represented by the boxcar function.
Because of the normalization of the functions, the integral is unity, thus the equation reduces to \eq{eq:XS}, thus the beam energy spread factors out from the analysis.
The only difference is the range of validity, which in the latter case requires the cross section to be slightly varying within the energy range covered by the non negligible part of the convolution in \eq{eq:XS_int}, instead of only in the energy range covered by the target.

If the slightly varying criteria is met, the derivation of the cross section from the yield is straightforward. However, if a sharp\footnote{The width of the resonance is significantly smaller than the target thickness.} resonance is present in the cross-section function in the relevant energy range, it alters the yield, while the above simple definition of $E_\mathrm{eff.}$ also becomes invalid. In such a case, $E_\mathrm{eff.}$ shall be determined numerically after considering an appropriate energy dependent cross section function, or employing an iterative approach \cite{Lemut08-EPJA}.
 
Besides the disadvantage of a more complicated data analysis in case of the presence of a resonance, an experiment with a beam of broader energy distribution has its advantage. By mapping the excitation function with overlapping energy distributions, sharp resonances cannot be missed, which can be the case with sharp beam energy distribution and data points further separated in energy than the target thickness. If a broader energy range of the excitation function is studied, this latter can be an issue.

\subsection{\label{sec:our}Energy uncertainties with gas-cells}

As mentioned in the previous section, the main difference between a solid target and a gas-cell measurement is the beam energy loss and straggling in the entrance foil.
Using charged particle stopping power tables, the beam energy loss can be calculated in a foil of known material and thickness. With more sophisticated charged particle transport codes like SRIM \cite{Ziegler10-NIMB} not only the energy loss, but the energy and position straggling can also be determined employing Monte Carlo calculations.

The uncertainty of the energy loss can be decomposed into two parts, first the foil thickness uncertainty and then that of the stopping power. These two shall be treated separately, since the first contributes to the point-by-point (p.b.p.) energy uncertainty, while the latter to the systematic (syst.) energy uncertainty.

In case of the p.b.p energy uncertainty in addition to the initial beam energy uncertainty, the entrance foil thickness uncertainty has its contribution.
Even if the same foil is used for several measurements, those can be threated as independent, since the slight inhomogeneity of the entrance foil is considered in its thickness uncertainty, thus a slightly different beam spot position may result in a slight change of the energy loss in the order of that uncertainty. The thickness precision quoted by the manufacturer is usually in the order of 10\%. Based on this, the p.b.p reaction energy uncertainty is about one tenth of the beam energy loss in the entrance foil. If higher precision is needed, the foil thickness shall be determined with the appropriate precision. 

The other part of the energy loss uncertainty is the stopping power uncertainty. This systematic uncertainty may cause the shift of the energies of the whole dataset together. The stopping power uncertainty is usually in the range of 3-10\% \cite{Ziegler10-NIMB} depending on the foil material, thus the syst. reaction energy uncertainty is less than one tenth of the beam energy loss in the entrance foil.

In principle, the beam energy loss in the gas (and not only in the entrance foil) would also have a point-by-point and a systematic uncertainty on the reaction energy. However, the overall small energy loss in the gas (few tens of keV) compared to the initial beam energy (few MeV) renders it negligible.

\section{\label{sec:cells}Development of the gas-cell targets and their application for activation experiments}

\begin{figure}[!b]
\center
\includegraphics[width=0.95\columnwidth]{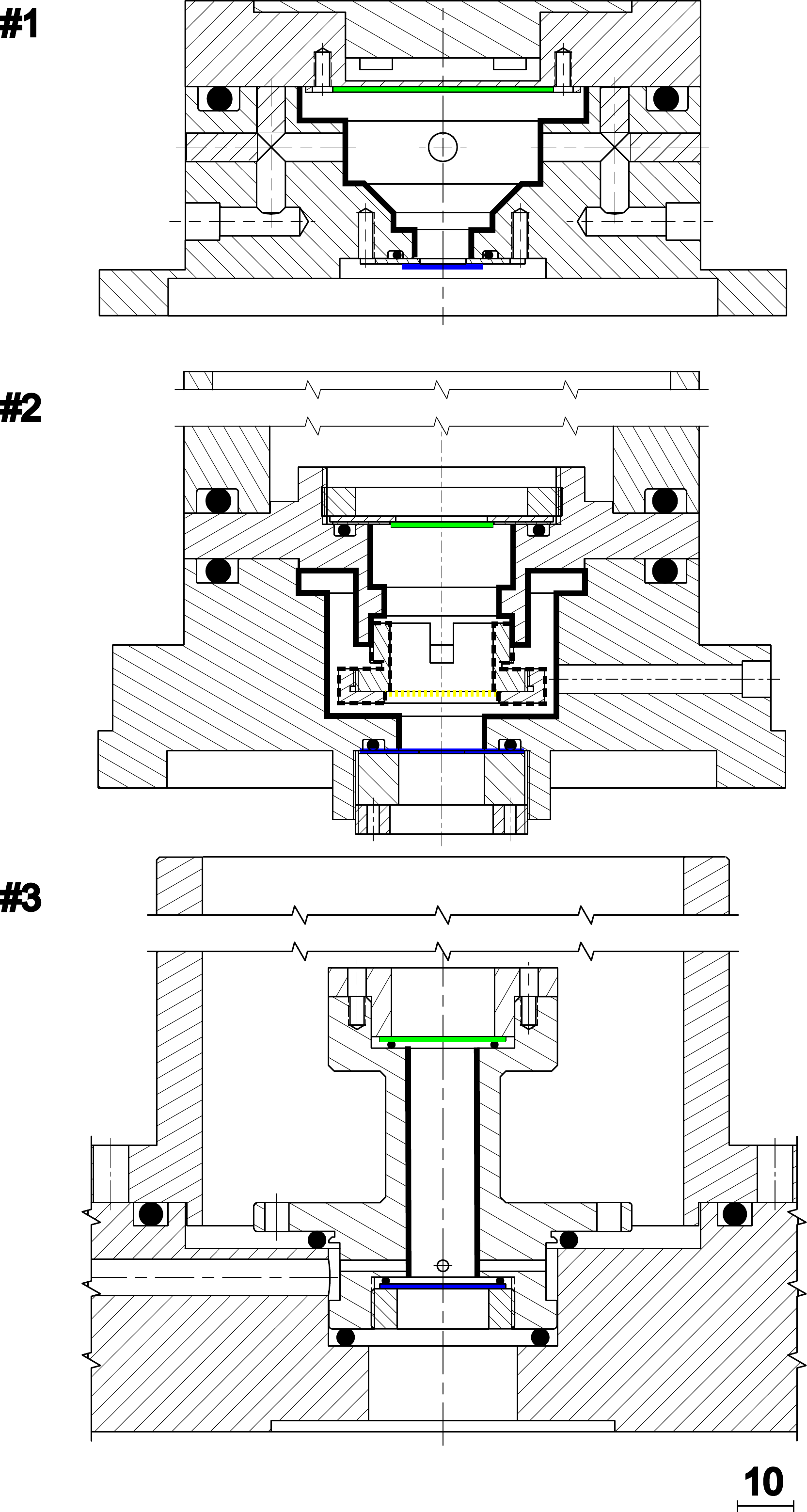}
\caption{\label{fig:cells} Technical drawings of the gas-cells on the same scale with a 10 mm mark, in the configurations as they were used for \HeHe investigations. Black filled circles are the cross section of the o-rings. 
The entrance window and catchers are shown in blue and green, respectively. The internal surface is highlighted by thicker black line. In case of \#2 an optionally placed internal catcher is marked by dashed yellow, and its holder by dashed black lines.
The beam enters in each cases from the bottom along the symmetry axes. In case of \#2, \#3 the vacuum closing top half is shown with broken view.}
\end{figure}

In recent years, several versions of gas-cells were designed, tested and used in nuclear astrophysics experiments in Atomki \cite{Bordeanu12-NIMA,Toth23-PRC,Szucs19-PRC,Bordeanu13-NPA,Halasz16-PRC,Szegedi19-NPA,Kovacs25-NIMA} (see \fig{fig:cells}). The driving force behind the continuous development was to eliminate the effects influencing the uncertainty of the experiments: the beam induced ones such as the heating effect of the beam and the surface desorption of the cell's inside walls contaminating the gas, and the possible permeation of the external air into the gas volume during the experiment. 
 In the forthcoming section we present the design steps, and compare their capabilities. 

\subsection{\label{sec:first}One windowed gas-cell (\#1)}

The gas-cell design \#1 had the simplest construction. This cell -- serving as basis of the following versions -- had a one windowed-configuration with a 1-$\mu$m-thick nickel foil glued on a stainless steel frame with an 8-mm-diameter opening for beam entrance. The outer part of the frame was pushing against an o-ring, sealing the gas volume from the beam-line vacuum. In this case the catcher (used to collect the radioactive reaction products) was a thick Oxygen-Free High Conductivity copper plate directly \linebreak mounted onto the water-cooled beam stop inside the gas-cell volume. The material of the catcher was chosen to have high thermal conductivity and high chemical purity. In case of low energy studies, it is crucial to minimize even the trace amount of impurities, on which parasitic reactions can produce unwanted activity, obscuring the signal of interest.  The catcher could be replaced after the irradiations by removing the complete beam-stop. For the separation of the cell volume from the environmental atmospheric pressure, o-ring sealings were used altogether about 450-mm nominal length. As the beam totally stopped in the copper plate being in thermal contact with the gas, its total energy was dissipated inside the gas-volume (see \fig{fig:cells} top).
With this gas-cell configuration the cross section of \HeHe was successfully measured in the beam energy range of $E_{\alpha} = 4.0-6.3$~MeV using about 300~mbar target gas pressure corresponding to about $2 \cdot 10^{19}$ atoms/cm$^2$ causing a typical 150 keV energetic target thickness \cite{Bordeanu13-NPA}.

The same cell design but with different construction materials was used for $^{124}$Xe irradiations with higher beam energies ($E_{\alpha} = 11-15$~MeV). 
At the time of the experiment, the previously used thin vacuum tight nickel foil commercially was not any more available. It was replaced by the thinnest available vacuum tight Al foils (6.5-$\mu$m thick) which caused twice as much beam energy loss as the \mbox{1-$\mu$m~Ni} foil, however, it was still acceptable because of the higher beam energies.
The copper catcher was not suitable any more, since longer lived (more than a few minutes in this case) activity could have been created in it via ($\alpha$,n) reactions.
A thinner catcher of 20-$\mu$m Al was chosen, because $\alpha$ irradiation does not create longer lived activity in aluminium. In addition, this thickness made it possible to deposit most of the beam power directly in the tantalum lined beam stop outside of the cell volume. However, still about 20\% of the beam power heated the catcher foil. 
The lower mass number of the catcher facilitated also the implantation of the reaction products. Because of the heavy target nuclei, the recoils from the reaction had only $300-500$-keV kinetic energy. 
Simulations were carried out using GEANT4 \cite{Agostinelli03-NIMA,Allison06-ITNS,Allison16-NIMA} with default physics lists to determine the implantation probability for such low energy nuclei. These simulations also include the energy loss of the created recoils in the gas volume, as well as the possible backscattering from the catcher. 
Their emergence points were uniformly distributed in the 50~mbar $^{124}$Xe gas volume, which corresponded to about $4 \cdot 10^{18}$~atoms/cm$^2$ target thickness. As a result, only those recoils were implanted into the catcher which were created in the last $1-2$~mm of the target, the others were stopped by the gas. The implantation depth was only $<150$~nm, which may also resulted in collection loss, if the already implanted atoms diffuse out.

\begin{figure}[t]
\center
\includegraphics[width=0.95\columnwidth, angle=0]{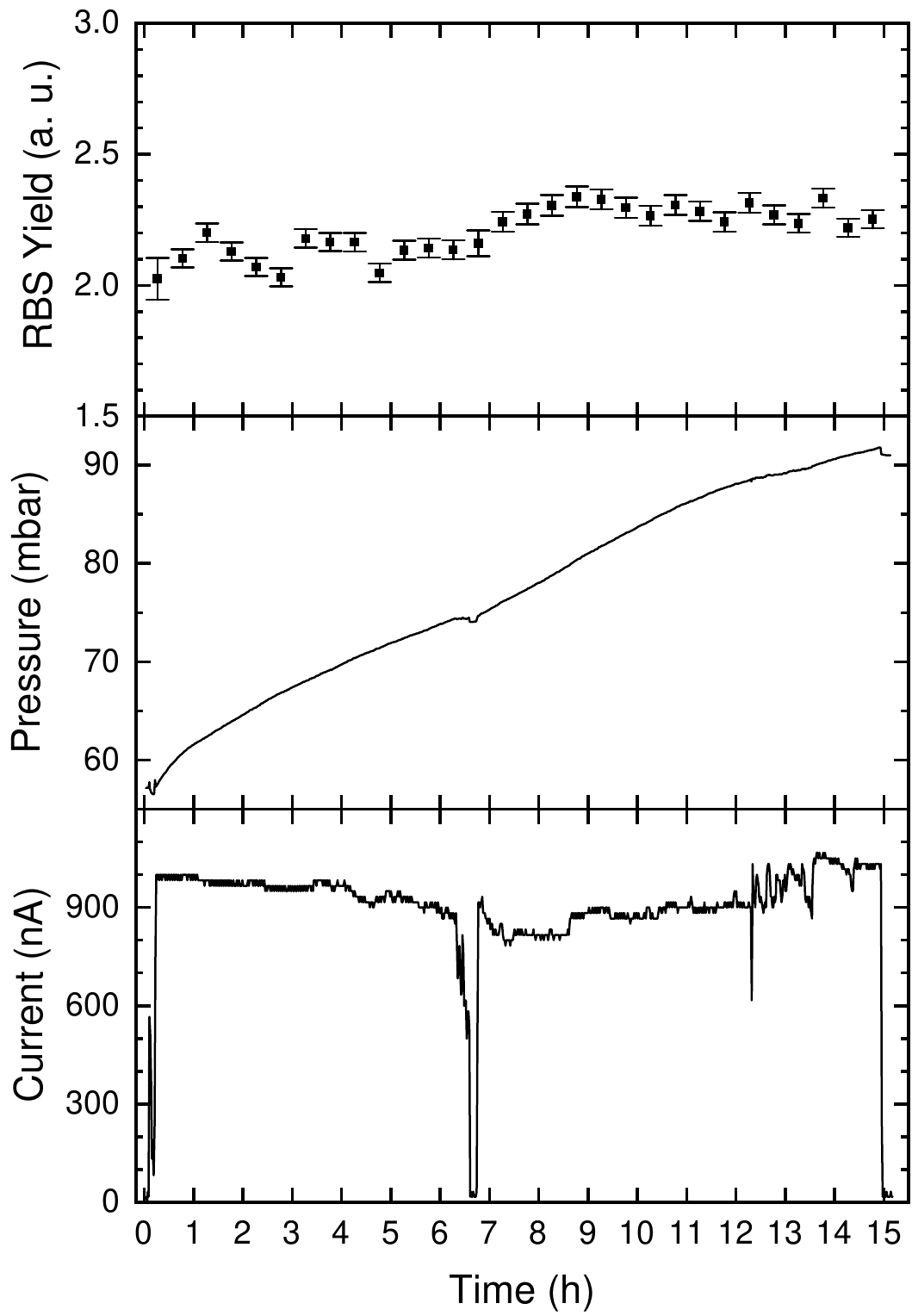}
\caption{\label{fig:press_vs_RBS}In the top panel the yield of the backscattered $\alpha$ particles from the Xe gas inside the gas-cell is shown. The middle panel presents the observed pressure, while the bottom panel shows the beam current. See text for discussion.}
\end{figure}

During the irradiations constant pressure increase was observed. A typical case is shown in \fig{fig:press_vs_RBS}, where the initial $\sim$55 mbar pressure increased to $\sim$90 mbar during the 15-h-long irradiation.
For the in-situ monitoring of the target nuclei a particle detector was placed in the irradiation chamber at 165$^\circ$ with respect to the beam direction in front of the gas-cell, detecting the backscattered projectiles from the entrance foil and from the gas (if that was kinetically possible). The backscattering yield from the Xe gas was measured and found to be constant despite the pressure increase (see top panel of \fig{fig:press_vs_RBS}). This confirms that there is no target material loss.
Although, the number of target atoms was proved to be constant, the possible reaction product implantation difficulties caused by the originally high pressure and the substantial pressure increase called for a new target design for the measurements on heavier targets.

This original one window configuration is perfectly usable, when the target pressure can be high due to the low mass target, and the reaction products reach and implanted into the catcher, which was the case for the \HeHe investigations \cite{Bordeanu13-NPA}. However, when more restricted pressure parameters have to be used, the \#1 version has a few design flaws.

\subsection{\label{sec:second}Two-windowed gas-cell  (\#2)}

The goal of the upgraded gas-cell design \#2 was to remove the bulk of the beam power deposition from the gas volume. The heating effect of the incident beam has two major drawbacks: firstly it has a direct impact on the gas-pressure. An increasing pressure  may mask a target material loss due to a small leak, and secondly as an indirect effect it increases the temperature of the wall of the cell which increases the outgassing of the surface. It is important to note, that the out-gassing rate follows the Arrhenius-law meaning it is highly sensitive to temperature.
The gas-cell target design presented in this section had a two window configuration with total length of 42~mm with the option to place a catcher foil inside the gas volume at a variable distance from the entrance window (see in \fig{fig:cells} middle).
As before, the thinnest commercially available vacuum tight Al foil of 6.5-$\mu$m thickness was used as entrance window.
The foil was directly pressed against a small o-ring with a stainless steel ring having a 12-mm-diameter opening. The bigger opening made the beam focusing easier, and the glueless solution made the foil exchange easier.
The exit foil was \mbox{10-$\mu$m-thick} aluminium glued onto a frame with \mbox{14-mm-diameter} opening.
The foil frame was pressed against an o-ring to close the gas-cell volume.
The beam stop and the rear part of the gas-cell, together with the catcher and exit foils can be detached from the beamline for easy foil exchange. When closed, an o-ring separated the internal volume from the outside atmosphere with a 450-mm nominal length.
This configuration was successfully used to measure $\alpha$-induced reactions on $^{124}$Xe. The catcher was placed 10 mm behind the entrance window, and only 5~mbar initial gas pressure before the irradiation was used, thus the target thickness was reduced to about $1.2 \cdot 10^{17}$ atoms/cm$^2$. With this configuration the recoils had enough energy to reach and implant into the catcher foil. The results were published in \Rref{Halasz16-PRC}.  

After the xenon campaign, the gas-cell was used to study the \HeHe reaction at energies around the $^7$Be proton separation threshold with a slight modification in its configuration. Because of the reaction kinematics, a much higher target density could be used without risking implantation loss due to the reaction products not reaching the catcher. The pressure was chosen to be 100~mbar, while no intermediate catcher foil was used, exploiting the 42-mm total length of the gas-cell, thus the closing foil served as the catcher. This combination resulted in a target thickness of about \mbox{$10^{19}$ atoms/cm$^2$}. Because of the higher pressures, a thicker entrance foil of 10 $\mu$m of Al was employed. The results of these measurement were published in \Rref{Szucs19-PRC}.
 
Subsequently, a new experiment was planned to measure the radiative proton capture reaction on  $^{124}$Xe.
Even though the half-life of the reaction product is suitable for activation studies, owing to the even lower projectile mass, the reaction kinematics posed a complication.
The distance of the catcher was set to 4.2 mm from the entrance foil and an even lower, 3 mbar initial gas pressure (corresponding to about $3 \cdot 10^{16}$ atoms/cm$^2$) was set to have reasonable implantation probability even at the lowest beam energy of 3 MeV. Successful implantation was achieved, and the half-life of the reaction product and its daughter isotope was precisely measured \cite{Szegedi19-NPA}.

In summary, the first two-window configuration was successfully employed both with low mass and heavy mass target gas studies, however, the still observable pressure increase requested further investigations (see \Sec{sec:pressure}) to mitigate systematic uncertainties. 
To reduce the possible permeation rate of air from the outside atmosphere and mitigate the effect of the out-gassing of the cell walls, a new gas-cell design was made.

\subsection{\label{sec:third}Two-windowed gas-cell surrounded by vacuum (\#3)}

The last gas-cell design \#3 was made with more favorable surface to volume ratio and o-ring length (see \fig{fig:cells} bottom). 
In this most recent design, a standalone replaceable gas-cell with selectable length can be placed into the gas handling system completely surrounded by the beamline vacuum.
Both the entrance and exit foils are pressed against o-rings with tantalum rings of 12-mm-diameter openings. There are two o-rings separating the internal gas-cell volume from the beamline vacuum.
The only possible way for air to enter the gas-cell is through the connections of the gas handling system, however, those are metal gasket sealed Swagelok VCR connectors, where the leakage rate has a $4 \cdot 10^{-11}$ std cm$^3$/s nominal value \cite{Swagelok-VCR}.
 With this design the air-flow turns backward: the pressure change in the gas-cell is dominated by the out-flow through the o-ring sealing into the vacuum of the beamline. However, as the permeation is proportional to the pressure difference, this rate is almost undetectable. This compact design also reduces the internal surface area of the cell by 20-50\% (depending on the size of the cell), further reducing possible surface desorption.

Three gas-cells were made with different lengths for several different reaction measurements, without an additional catcher foil inside the gas volume. The exit foil always acts as a catcher.
One version with about 42-mm length was used to study the \HeHe reaction at higher energies of $E_{\alpha} = 11.0-20.0$~MeV \cite{Toth23-PRC}. In this case an even thicker catcher foil (up to 25~$\mu$m) was needed to be able to collect the $^7$Be recoils with higher energy, which would penetrate the standard 10~$\mu$m aluminium. 
A gas-cell with an intermediate length of 10~mm was used to study the $^{86}$Kr($\alpha$,n)$^{89}$Sr reaction \cite{Kovacs25-NIMA}. Since this target gas is lighter than $^{124}$Xe, 13~mbar initial target pressure was chosen (corresponding to a target thickness of about $3 \cdot 10^{17}$ atoms/cm$^2$), and not allowed to increase above 20~mbar to get at least 80\% implantation probability. This is an arbitrarily chosen value, based on the maximum allowed additional uncertainty to the derived reaction cross section. Assigning a conservative 15\% uncertainty to the simulated particle transport, a 20\% reaction product loss causes a maximum 3\% extra uncertainty in the cross section.
In this case, the reaction cross section was also determined at selected energies employing solid foil targets implanted with $^{86}$Kr \cite{Kiss25-AA}. These gave consistent results, supporting the accurate recoil implantation probability calculations in case of the gas-cell target.

Further tests for gas-cell optimization are ongoing. With the 10-mm long cell, the $^{124}$Xe$+\alpha$ reaction is planned to be investigated again, while a new 3-mm long gas-cell is planned to be used for the $^{124}$Xe$+$p.

\subsection{\label{sec:pressure} Investigation of the pressure increase in the cells}

As seen in \fig{fig:press_vs_RBS}, a pressure increase of about 2.3 mbar/h is observed in case of the \#1 cell when it experienced about 1 $\mu$A beam current.
If that were a pure temperature effect, the gas temperature should have risen about 190 degrees, which is impossible given the water cooled body of the complete cell. The cooling water temperature was always around 24~$^\circ$C.
The temperature effect is the small instantaneous pressure change of about 0.5 mbar occurred when the beam was switched on or off. This corresponds to a temperature change of less than 3~$^\circ$C. 
The beam constantly deposits energy inside the gas volume, while the gas is cooled by the approximately constant temperature gas-cell walls. An equilibrium temperature is quickly achieved after the power source (the beam energy deposition) changes.
This increase or decrease can be clearly seen at the beginning where after a two minute irradiation the beam was stopped for a few minutes and restarted again. Then around the 7$^\mathrm{th}$ hour, the beam was lost again. Since in that case the current decreased continuously within half an hour, the down-jump is not prominent. However, when the beam was instantly regained, the up jump is visible. Finally, switching off the beam caused a down jump.
\begin{figure}[t]
\center
\includegraphics[width=0.95\columnwidth]{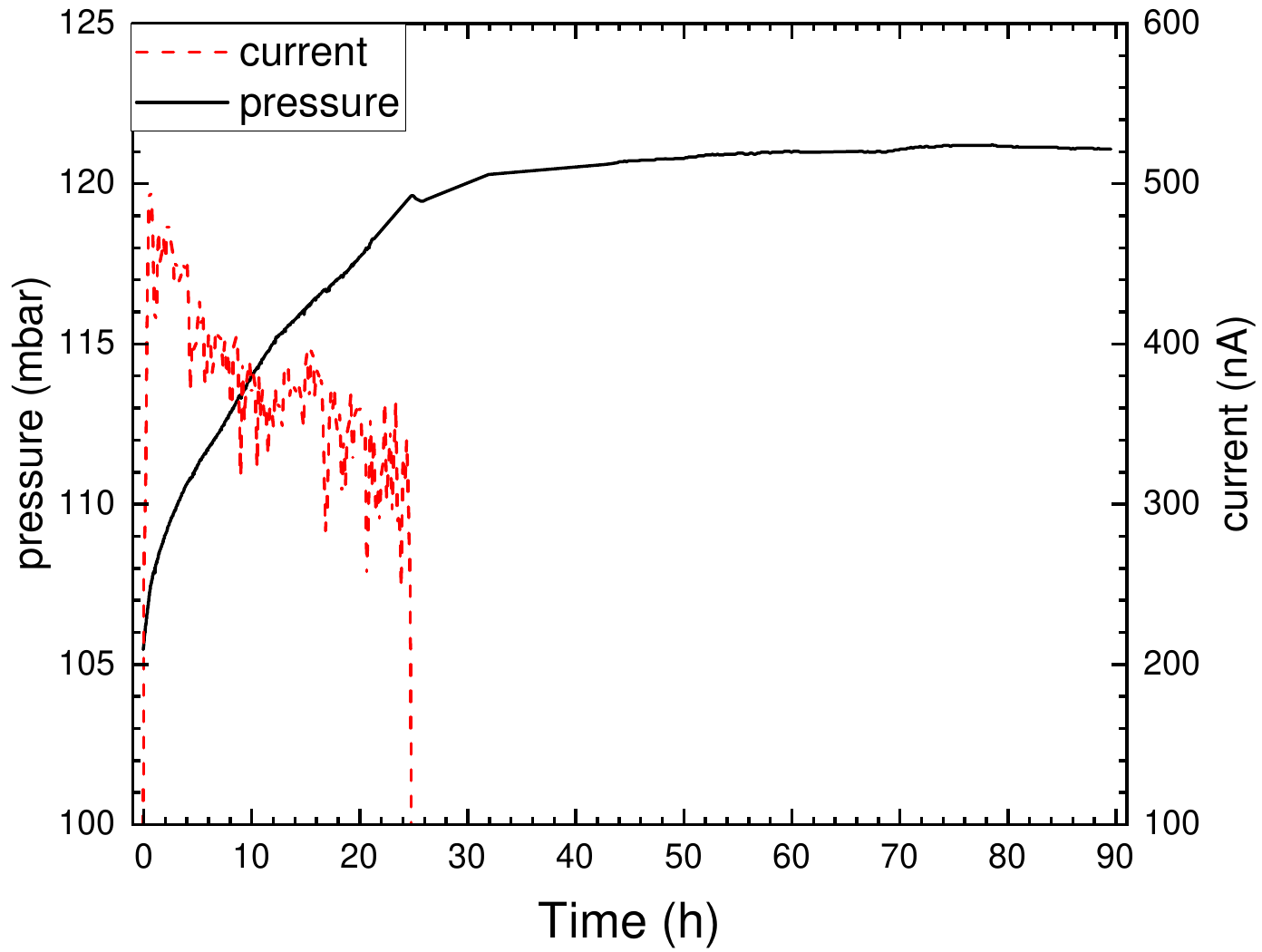}
\caption{\label{fig:pressure}Typical pressure increase during an irradiation and subsequent no-beam pressure variation in case of the \HeHe campaign. Substantial pressure increase is visible during the irradiation, while the later pressure variations are compatible with the effect of the room temperature changes.}
\end{figure}

In case of the \#2 cell, removing the bulk of the power deposition from the cell internal volume, a continuous pressure increase was still observable during the irradiation, however, with lower overall increase rate of about 0.6~mbar/h.
The pressure was recorded for longer time after a \HeHe measurement to further investigate the pressure development. As can be seen in \fig{fig:pressure} the added pressure rise did not vanish for days after the irradiation, while the observed pressure variation after the beam was stopped is compatible with room temperature changes of few $^\circ$C. This strengthens the implausibility of the above discussed high temperature change during the irradiation, and points towards possible air in-leaking from the atmosphere.

\begin{figure}[t]
\center
\includegraphics[width=0.95\columnwidth, angle=0]{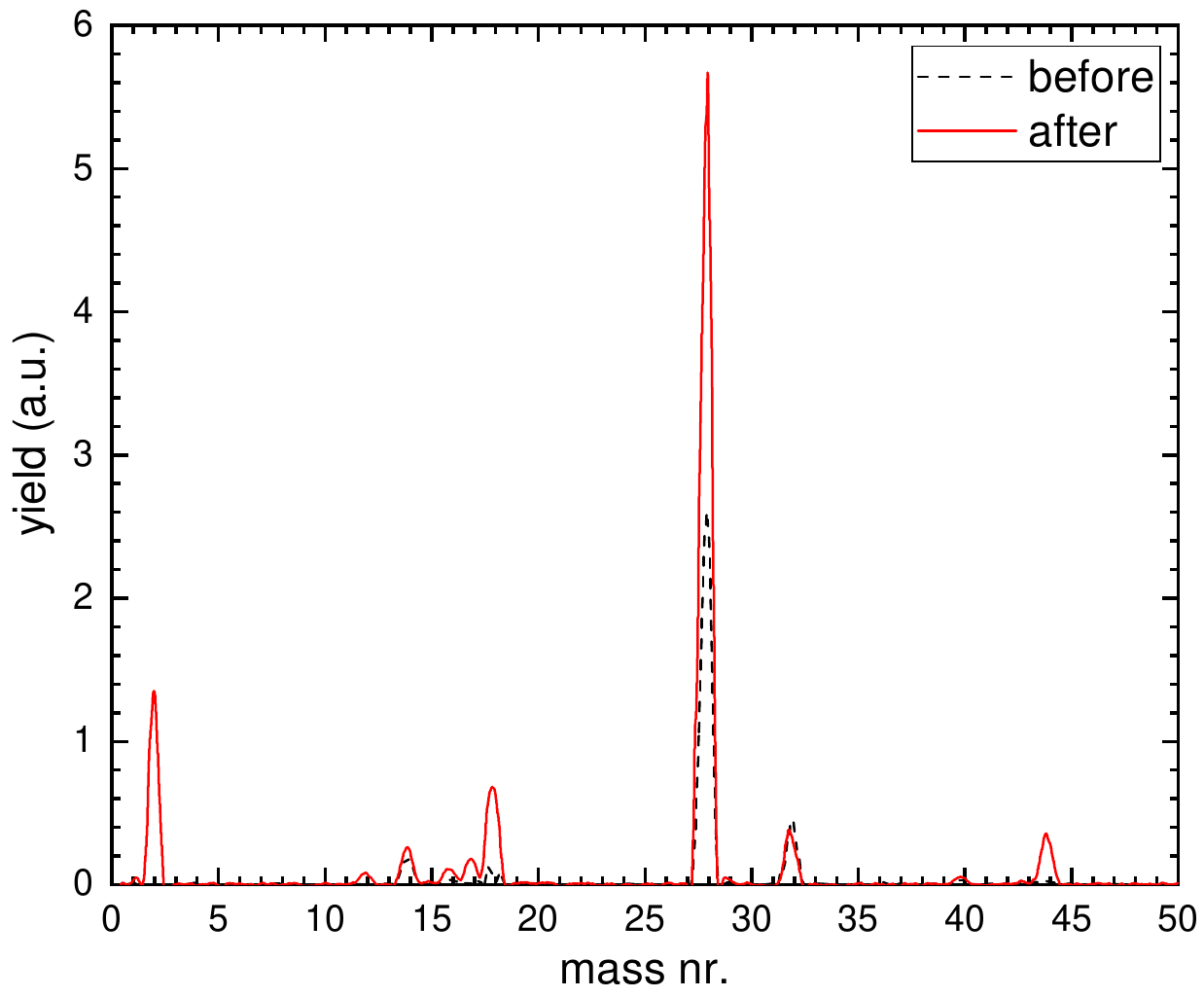}
\caption{\label{fig:comp}Mass spectrum of the gas sample from the cell before and after a 5.5-hour long irradiation of a 50 mbar natural Xe target. Only the low mass part of the spectrum is shown.}
\end{figure}

This was tested by connecting a residual gas mass spectrometer to the vacuum system, and a sample was drawn from the cell before and after the irradiation of natural xenon (see \fig{fig:comp}), while the complete irradiation chamber was placed into a balloon filled with argon gas.
Only qualitative analysis could have been done, since no ionization correction was employed. Thus the number of counts in the peaks are only proportional to the amount of the given gas, but those cannot be compared to each other. Before the irradiation some nitrogen, oxygen, water vapor and argon is already present (mass nr. 14, 28 for N, N$_2$; 16, 32 for \mbox{ O, O$_2$;} 18 for H$_2$O, and 40 for Ar, respectively, where the single ions are formed during the ionization process in the measuring device). These are most probably from the vacuum line between the gas-cell and mass spectrometer (which was used previously for argon measurements). After the irradiation, appearance of mass peaks at 2, 12, 17, 44 was observed in addition to the increase in nitrogen and water vapor signals. The new peaks correspond to H$_2$, C, HO and CO$_2$, respectively. Formers are molecular fragments of water and carbon-dioxide which are formed during the ionization process. A possible in-leaking is excluded, since no increase in mass 40 appeared, which is the mass of the surrounding argon gas.

\begin{figure}[b]
\center
\includegraphics[width=0.95\columnwidth, angle=0]{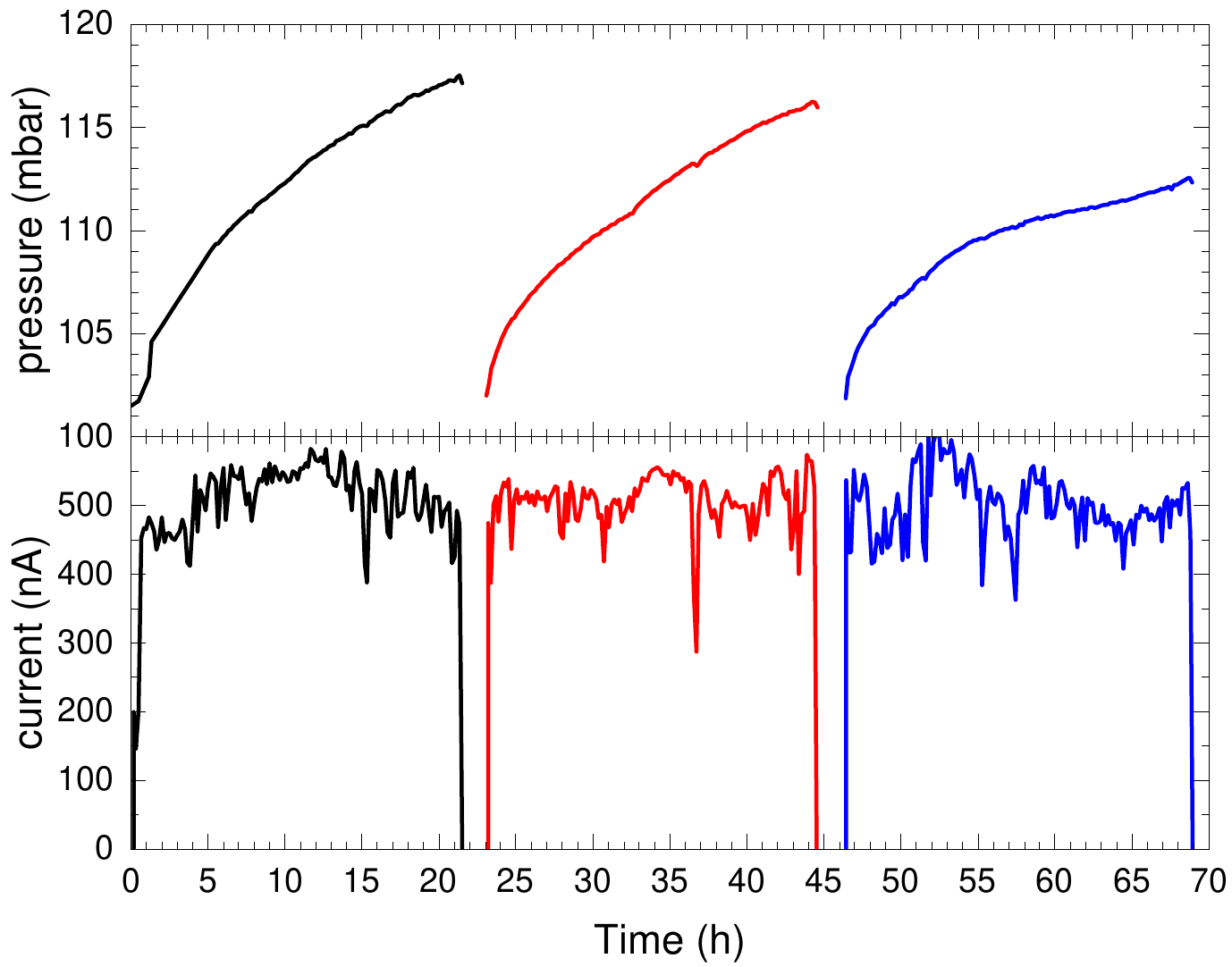}
\caption{\label{fig:press_change}Pressure development (top panel) and the beam current (bottom panel) in three subsequent irradiations. Between the irradiations only the exit foil was changed, and the cell was refilled with about 100 mbar gas. The final pressure value decreases, indicating less and less out-gassing. See text for details.}
\end{figure}

\begin{figure}[t]
\center
\includegraphics[width=0.95\columnwidth, angle=0]{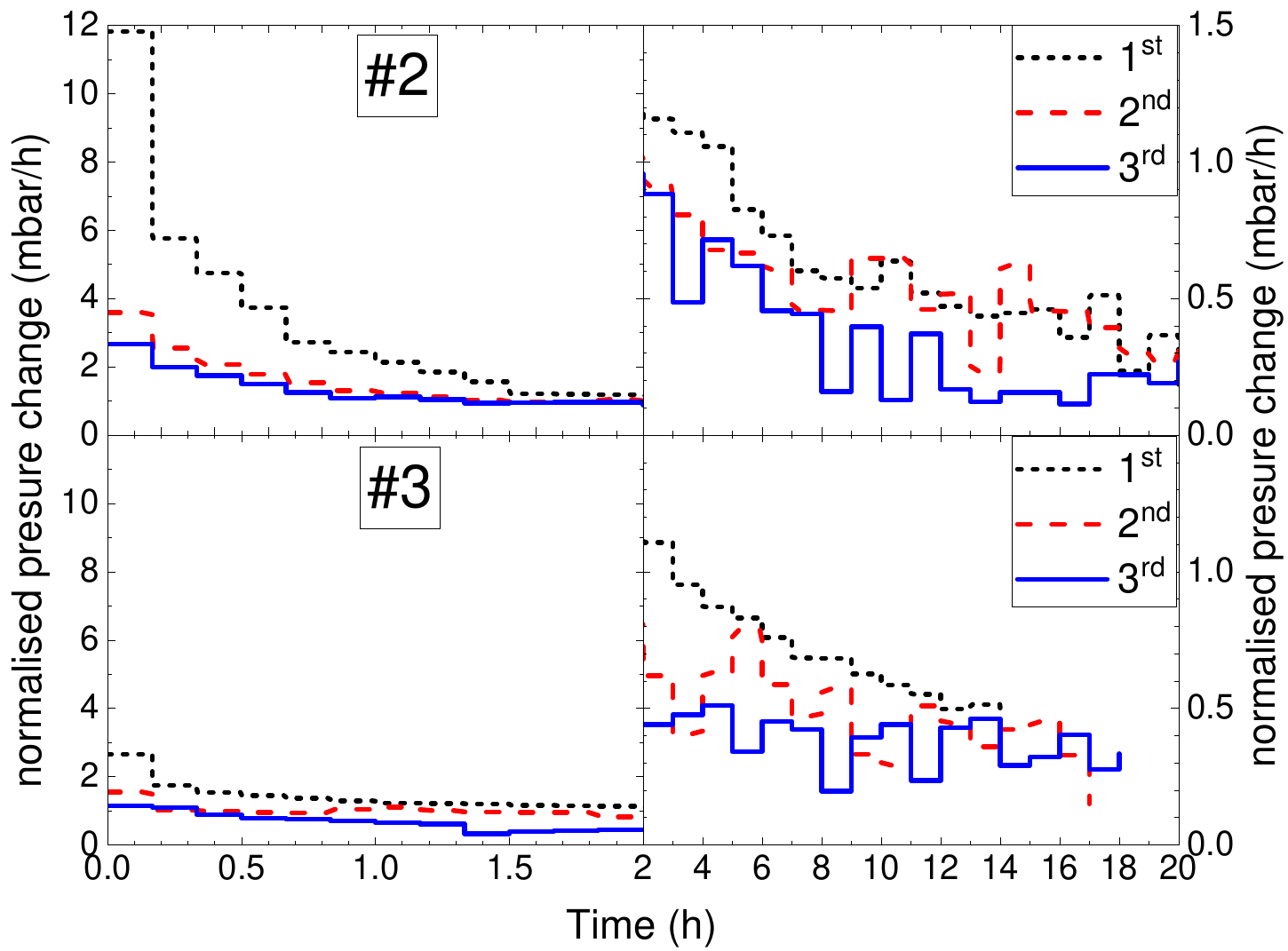}
\caption{\label{fig:press_norm} Current normalized pressure variation by the irradiation time for subsequent irradiations for cell \#2 (\#3) at the top (button), respectively. Note the scale change for both axes at the middle of the plots. For cell \#2 the same base data are used as in \fig{fig:press_change}. See text for details.}
\end{figure}

Since the extra gas contains elements and molecules typically found in the atmosphere, but no in-leaking is observed, it was concluded that the cause of the pressure rise is the out-gassing of the foils and gas-cell walls, which were exposed to air before the irradiations.
This hypothesis is supported by the observation that a lower pressure increase rate is observed for subsequent irradiations (see \fig{fig:press_change}).
Another feature is also visible in \fig{fig:press_norm}, namely the pressure increase rate is higher at the beginning of the irradiation and then settles to an approximately constant rate.

The \#3 cell design intended to eliminate the possibility of air in-leaking, and decrease the out-gassing by reducing the internal surfaces.
A similar investigation that plotted in \fig{fig:press_change} has been done with the new cell, and the observed pressure change rate was similar to that of cell \#2.
Since the out-gassing rate is temperature dependent, and a higher beam current results in higher power deposition in the foils and gas-cell, thus slightly higher equilibrium temperatures, the pressure increase has to be normalized to the beam current.
This was done by dividing the actual pressure increase by the average beam current within the same 10 minutes and scaled to a current of 500 nA. The resulting curves are presented in \fig{fig:press_norm}. A much higher increase rate is observed in the first $1.5-2$ hours in each case settling to below 0.5~mbar/h. There is still a decrease of the rate, but with much lower slope. The initial hight rate is higher for the first irradiations of the weeks and the lowest for the lasts.
The newest design has a lower initial rate, then in the latter phases the curves from the two cells run parallel.
The first phase can be attributed to the out-gassing of the entrance and catcher foils together with the cell walls, later the absorbed gas is removed from the intermediate surfaces, and the rest out-gassing comes from deeper layers of the surfaces.
During the short venting of the cell for the catcher foil exchange much less air is absorbed, which causes the lower increase rate at the begging of the second and third irradiations, however, from the deeper layers the out-gassing rate stays the same.
The advantage of the \#3 cell is the smaller immediate pressure increase, due to its smaller internal surfaces.

A gas-cell heating is planned in future experiments before the irradiations to remove as much absorbed gas from the inner surfaces as possible to decrease the pressure increase rate.
This will be especially useful in case of the heavier target experiments especially in the case of (p,$\gamma$) reactions, where too much gas may hinder the implantation of the slow moving reaction product into the catcher.

\section{\label{sec:hehe}The new \HeHe investigation}

\subsection{\label{sec:exp}Experimental details}

As an example of the application of one of the above described gas-cells, the \HeHe reaction cross section measurement using gas-cell design \#2 will be presented in the following. The cell had nominally 10-$\mu$m thick aluminium foil entrance and exit windows. The measured entrance foil thicknesses were between $10.3 - 10.9$~$\mu$m determined  by alpha energy loss similarly to that in \Rref{Szucs19-PRC}. A given entrance foil was used for multiple irradiations. This thickness causes $1250-1500$~keV energy loss of the beam particles in the energy range of the present investigations ($E_\alpha = 7.5 - 8.5$ MeV) and would widen the energy distribution of an ideal mono-energetic beam via straggling to an approximately Gaussian form with a standard deviation of \mbox{$31-33$~keV}. This shall be quadratically added to the original beam energy distribution if it is not mono-energetic as in the present case.
The exact thickness of the exit foils was not relevant, because of the reaction kinematics the $^7$Be recoils have an energy in the range of $3.3-4.3$~MeV. This ensures that the $^7$Be is implanted at a depth of $4.9 - 6.1$~$\mu$m into the exit foil, well within the nominal 10-$\mu$m thickness. To minimize the parasitic activity production on any trace amount of foil impurities, a high purity (99.999\% Al)\footnote{Goodfellow AL00-FL-000175} foil was chosen as exit window. 
The length of the gas-cell between the two foils was $42\pm1$~mm and enclosed 99.95\% isotopically pure $^3$He gas with a pressure of about 100 mbar. The areal number density of the target was derived using the ideal gas law as mentioned before.
These parameters resulted in an energetic target thickness of about $\Delta$E$_\alpha =56-62$~keV in the investigated energy range.
With the typical beam currents, the local thinning of the gas (the so-called, beam heating effect) is in the order of maximum 1\% using the formula from \Rref{Marta06-NIMA}. This correction was taken into account in the cross section derivation.

The doubly charged $\alpha$ beam was provided by the \linebreak \mbox{MGC-20} cyclotron of Atomki, where the gas-cell target was installed at the 2A beamline \cite{Biri21-EPJP}. The typical length of the irradiations varied between $19-24$~h with a typical beam current of $0.3-0.5$~$\mu$A. After the irradiations, the gas was released and the exit foil replaced before a new run was started.

The accumulated $^7$Be activity in the exit foil was then determined by \g~spectroscopy utilizing the 477.6-keV \g~ray following 10.44\% of the $^7$Be decays \cite{Tilley02-NPA}. A high purity germanium detector with 100\% relative efficiency was used (the same as in \Rref{Szucs19-PRC}), which was placed into a \mbox{10-cm} thick commercial lead shielding layered by cadmium and copper to further reduce the low energy background. The absolute detection efficiency was determined with calibrated radioactive sources, similarly as described in Refs. \cite{Toth23-PRC,Szucs19-PRC}. The activity of each sample was determined at least 3 different times to follow its decrease, which was always consistent with expectations considering the half-life of $^7$Be. At the end the results from the different countings on the same sample were combined to reduce the statistical uncertainty. 

\subsection{\label{sec:res}Results and discussion}

\begin{figure}[b]
\center
\includegraphics[width=0.95\columnwidth]{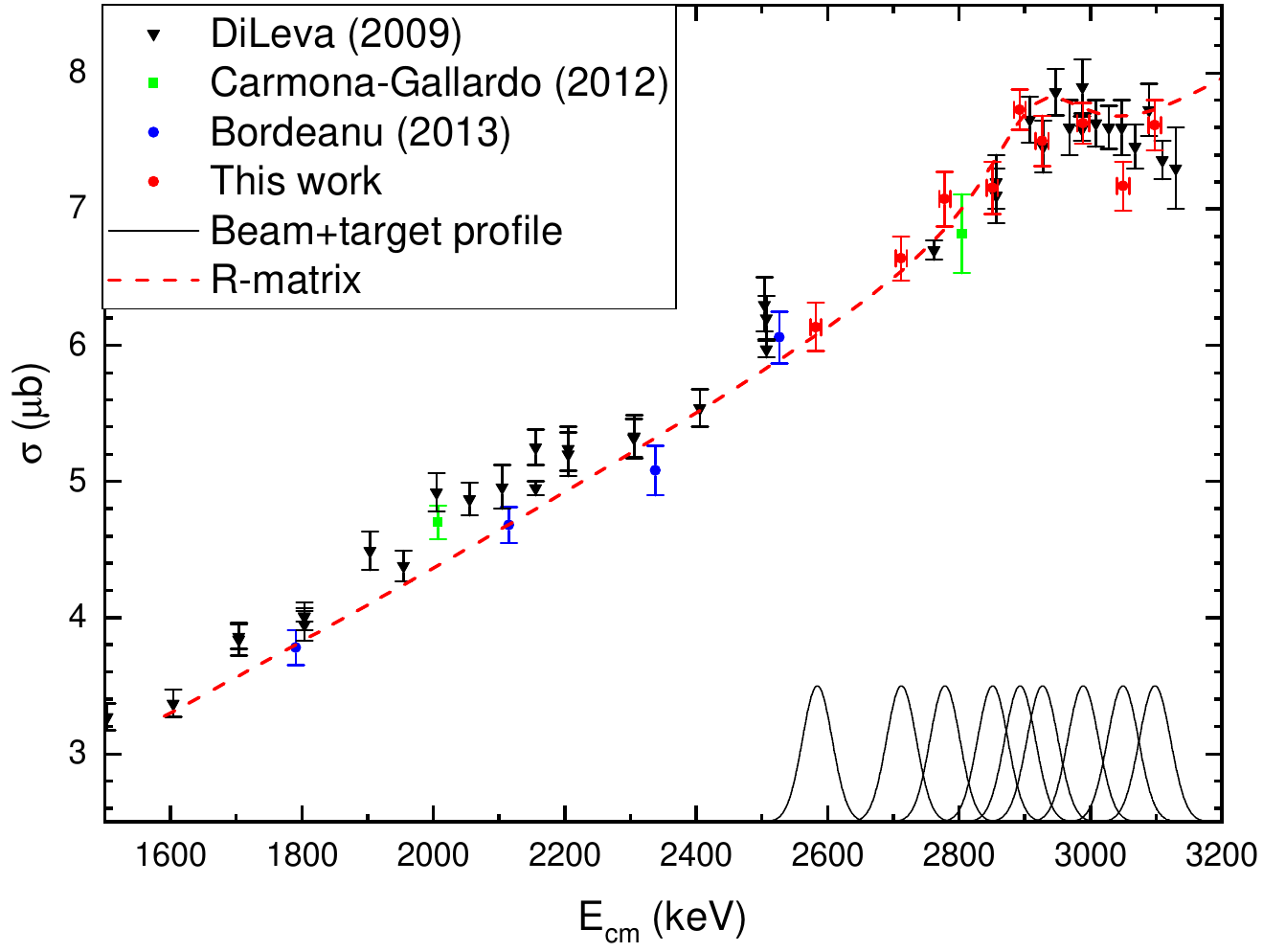}
\caption{\label{fig:res}Experimental \HeHe results compared to literature data from \Rref{Bordeanu13-NPA,DiLeva09-PRL,Carmona-Gallardo12-PRC}. Only statistical uncertainties are shown. An additional 4.6\% scaling uncertainty shall be considered for the present dataset. The horizontal error bars show the point-by-point energy uncertainty only, while the systematic energy uncertainty of the dataset is about 17~keV. The R-matrix fit using the parameters from \Rref{Toth23-PRC} is also displayed. On the bottom axis, the integrand for the measured points is shown in arbitrary scale. See text for details.}
\end{figure}

Cross section data have been measured up to the energy range of the first resonance in the \HeHe reaction with about 100-keV $\alpha$-energy steps corresponding to \mbox{$\approx$40-keV} steps in the center-of-mass (c.m.) system, and with bigger leaps below.
Additional to the statistical uncertainty of the \g-ray detection, a 4.6\% systematic energy independent common mode uncertainty had to be taken into account when the absolute scale of the new dataset is compared to other studies. This latter is the quadratic sum of the following partial uncertainties: charge integration (3\%), \g-detection efficiency (2.5\%), cell length (2.4\%) which is the dominating target thickness uncertainty, and \g-ray branching ratio (0.4\%). The implantation probability may cause an additional point-by-point uncertainty, however, in case of this light system, as mentioned before, the implantation depth is so high, that the implantation loss is negligible.
The resulting cross sections are shown in \fig{fig:res} and in \tab{tab:res}.
\begin{table*}[t]
\caption{The obtained cross section dataset for \HeHe with the statistical and systematic uncertainties, and with the corresponding effective center-of-mass energies with the point-by-point and systematic energy uncertainties. The standard deviation of the  convolution from \eq{eq:XS_int} estimated by a Gaussian is also listed, so are the total detection live times.}
\label{tab:res}
\center
\begin{tabular}{c c c c c }									
$E_{\alpha}$ 		&	$E_\mathrm{c.m.}^\mathrm{eff.}$\,$\pm$\,$\Delta$$E^\mathrm{eff.}_{\mathrm{c.m.}~\mathrm{p.b.p.}}$\,$\pm$\,$\Delta$$E^\mathrm{eff.}_{\mathrm{c.m.}~\mathrm{syst.}}$	&  $\mathrm{SD}_\mathrm{c.m.}^\mathrm{conv.}$ 	&	$\sigma$\,$\pm$\,$\Delta$$\sigma_\mathrm{stat.}$\,$\pm$\,$\Delta$$\sigma_\mathrm{syst.}$ 	& t$_{total-live}$\\
 (MeV)				&	(keV)	&	(keV)		&	($\mu$barn)  & (h) \\
\hline
7.50	&	2582\,$\pm$\,10\,$\pm$\,18	& 18 &	6.13\,$\pm$\,0.18\,$\pm$\,0.28	& 421.0\\
7.70	&	2712\,$\pm$\,10\,$\pm$\,18	& 18 &	6.64\,$\pm$\,0.16\,$\pm$\,0.31	& 220.5\\
7.85	&	2778\,$\pm$\,10\,$\pm$\,17	& 18 &	7.07\,$\pm$\,0.20\,$\pm$\,0.33	& 305.2\\
8.00	&	2851\,$\pm$\,10\,$\pm$\,17	& 18 &	7.16\,$\pm$\,0.19\,$\pm$\,0.33	& 292.7\\
8.10	&	2892\,$\pm$\,11\,$\pm$\,17	& 19 &	7.73\,$\pm$\,0.15\,$\pm$\,0.36	& 436.7\\
8.20	&	2927\,$\pm$\,12\,$\pm$\,17	& 19 &	7.50\,$\pm$\,0.19\,$\pm$\,0.34	& 257.3\\
8.30	&	2989\,$\pm$\,11\,$\pm$\,17	& 19 &	7.63\,$\pm$\,0.15\,$\pm$\,0.35	& 453.1\\
8.40	&	3050\,$\pm$\,11\,$\pm$\,16	& 19 &	7.17\,$\pm$\,0.18\,$\pm$\,0.33	& 214.8\\
8.50	&	3098\,$\pm$\,11\,$\pm$\,16	& 19 &	7.62\,$\pm$\,0.19\,$\pm$\,0.35	& 231.5\\
\end{tabular}								
\end{table*}

The effective energies ($E_\mathrm{c.m.}^\mathrm{eff.}$) were determined considering the energy loss of the beam in the entrance foil and in the gas, setting it at the half of the energetic target thickness.
The original beam energy spread of 0.3\% is an inherent feature of the cyclotron accelerator, because the track separation at the deflector at the output of the accelerator is not enough to only extract a single energy. 
The beam energy uncertainty is taken also to be this value, which in turn is the initial point-by-point energy uncertainty. Before a given irradiation, the fields of the bending magnets are tuned to reach a suitable current at the target station, then those are fixed for the whole period of the irradiation. Any drift in the initial beam energy of the cyclotron causes a beam intensity loss (as one can see it e.g. in \fig{fig:press_change}), but not a change in the beam energy at the target.
Additional to this, as discussed in \Sec{sec:our}, the foil thickness uncertainty adds to the p.b.p. energy uncertainty. Since the energy loss is about $10-20$\% of the initial beam energy, this caused a maximum of $0.2$\% effect. Together with the initial beam energy uncertainty, the point-by-point energy uncertainty ($\Delta$$E^\mathrm{eff.}_{\mathrm{c.m.}~\mathrm{p.b.p.}}$) is estimated to be $0.35-0.4$\%.
The $\alpha$ stopping power uncertainty in aluminium is 3.5\% according to \Rref{Ziegler10-NIMB}. Since the thickness of the foil was measured by $\alpha$ energy loss, and the beam was also $\alpha$ particles, this systematic uncertainty was only once taken into account to avoid double counting. This affects systematically the energy of the whole dataset in the order of $0.35-0.70$\% ($\Delta$$E^\mathrm{eff.}_{\mathrm{c.m.}~\mathrm{syst.}}$) relative to the beam energy.

The initial beam has already a 0.3\% spread, and it passes through the entrance aluminium foil, which causes an additional beam spread due to straggling.
This combined effect causes an $\alpha$ beam energetic width when entering the active volume between $39-41$~keV, which is comparable with the target thickness of $56-62$~keV. The convolution of these two results in a function, which is still well described by a Gaussian with a standard deviation of $43-44$~keV. This value is converted into the c.m. system and called as $\mathrm{SD}_\mathrm{c.m.}^\mathrm{conv.}$ in \tab{tab:res}. Since this is one sigma of a Gaussian, the energy range where it is not negligible covers an interval of about 110 keV in the c.m system. To demonstrate this, the integrands are shown in \fig{fig:res}.

As seen in \fig{fig:res}, the new results are in very good agreement with the previous study \cite{DiLeva09-PRL}, and the position and width of the resonance seen by \Rref{Barnard64-NP} in the scattering channel, and by \Rref{DiLeva09-PRL} in the capture channel, is confirmed.
A limited R-matrix fit using the AZURE2 code \cite{Azuma10-PRC} was performed to evaluate the impact of the new dataset on the derived resonance parameters. A full R-matrix investigation is beyond the scope of the present article. Four slightly different R-matrix parameter set were utilized \cite{Toth23-PRC,Kontos13-PRC,deBoer14-PRC,Odell22-FP}. The resonance parameters (i.e. energetic position, $\alpha$ and $\gamma$ widths) were fitted using the present data, and those from \Rref{DiLeva09-PRL} which lie in the present energy range (i.e. $2.75-3.15$ MeV), while other parameters were kept fixed. No particle scattering data were used for the present fit. The scale of the two datasets was allowed to vary within their systematic uncertainties.
In \fig{fig:res} the best fit curve based on the most recent R-matrix parameters from \Rref{Toth23-PRC} is plotted, without scaling of the datasets. The fit itself resulted in a scale factor of 1.020 and 1.012 for the \cite{DiLeva09-PRL} and the present dataset, respectively, well within the common-mode uncertainty of those, 5\% and 4.6\%, respectively.
\begin{figure}[b]
\center
\includegraphics[width=0.95\columnwidth]{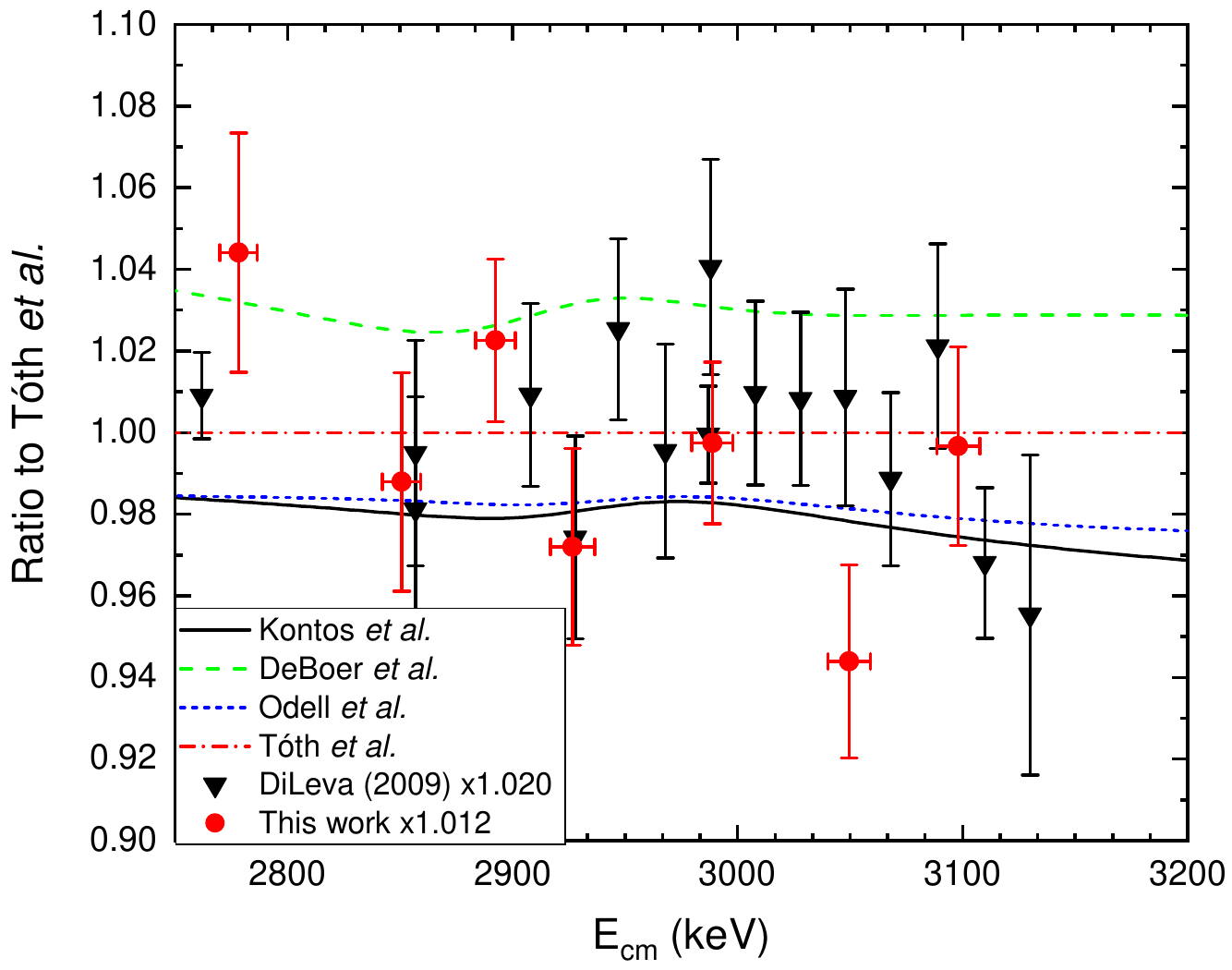}
\caption{\label{fig:ratio} The present data and those from \Rref{DiLeva09-PRL} and several R-matrix fits with parameters from \Rref{Kontos13-PRC,deBoer14-PRC,Odell22-FP} are compared with the fit using \Rref{Toth23-PRC} parameters in the energy range of the resonance. Only statistical uncertainties are shown. See text for details.}
\end{figure}

The ratio of the fits with the different parameter sets in the energy range of the resonance is presented in \fig{fig:ratio} together with the ratio of the scaled experimental points and the R-matrix fit.
In general all the fits reproduce the data within 4\%.
In the previous works, the energetic position and the alpha width of the resonance was constrained by scattering data. The position was 4.58~MeV, while the $\alpha$ width was about 160~keV, and the reported $\gamma$ widths were in the range of $36-50$~meV. If the scattering constraint is omitted using only the limited \cite{DiLeva09-PRL} data a slightly lower position of 4.54~MeV is obtained with consistent widths. Using only the present data the position became even lower, 4.49~MeV and higher widths are obtained. Combining both datasets, the present fit favors an energetic position of 4.52~MeV, and widths of  $\mathrm{\Gamma}_\alpha = 200$~keV and $\mathrm{\Gamma}_\gamma = 65-80$~meV.
For a more detailed analysis the point-by-point energy uncertainties shall also be taken into account, as well as the common mode energy uncertainty of about 17~keV, which may shift the whole set towards higher energies, resulting in a higher resonance position, consistent with that obtained from the scattering dataset.
In general, for a more precise resonance parameter determination, higher energetic precision dataset would be desirable.

\section{\label{sec:scatt}Gas-cell for particle scattering}

\begin{figure}[t]
\center
\includegraphics[width=0.95\columnwidth]{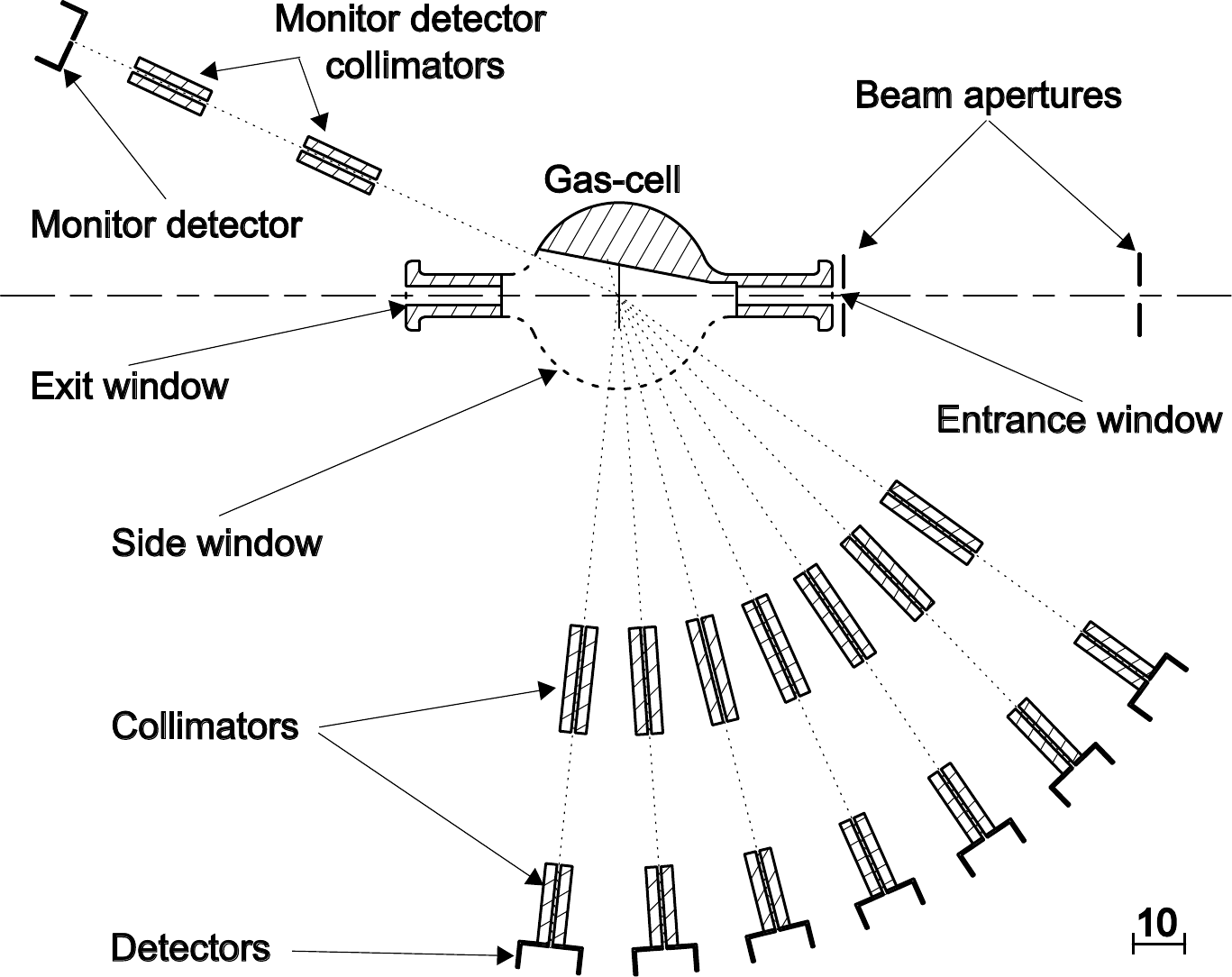}
\caption{\label{fig:scatt}Top view of the scattering gas-cell design with a 10~mm scale bar. The foil windows are marked with dashed lines, while the long detector collimators are marked with shadings. The detectors are sitting in the bracket holders. The beam enters from the right through the two marked apertures, and leaves the cell on the left toward a Faraday cup. The scattering directions are marked by dotted lines.}
\end{figure}

Besides the gas-cells used for activation, a cell for particle scattering experiments is also under development. Unlike in the previous case, the scattering experiment is less sensitive to pressure changes, the main point of the design is to find the balance between the precision desired in the experiment and the possible mechanical realization.
The design drawing of this scattering gas-cell together with the beam and detector collimators is shown in \fig{fig:scatt}.
In case of a particle scattering experiment, other considerations have to be taken into account than for the activation studies. For example, the beam particles can scatter on the entrance and/or exit foils not only on the gas, spoiling the recorded spectra.
However, with suitable collimation of both the incident beam and the detectors, these unwanted scattered particles can be avoided.
In our case two grooves of 3 cm and 4 cm in length 5 cm apart were in front of the detectors (see \fig{fig:scatt}). The width of the grooves was 1.5 mm for the three more forward detectors, and 2 mm for the four more backward detectors. The monitor detector had a similar collimator with 1 mm groove width. 
As beam collimator, two apertures with 4 mm circular openings 6 cm apart was placed in front of the gas-cell having a 5 mm entrance hole. This defined the beam position on the gas-cell.
With a collimated detector telescope arrangement, each of the detectors see only a particular extended slice of the gas volume, which is well defined by the collimator acceptance factor \cite{Silverstein59-NIM}. This so-called G factor are 0.3; 0.6; and 1.1 $\mu$m for the 1; 1.5; and 2 mm grooves, respectively. Dividing these with the length of the beam path, which a given detector sees (3~mm in the present case), the factors are $1 \cdot 10^{-4}$; $2 \cdot 10^{-4}$; and $ 3.7 \cdot 10^{-4}$, respectively, to be compared with opening angles of $1-2 \cdot 10^{-4}$ sr in case of  a typical solid target scattering experiment \cite{Kiss13-PRC}.  The entrance and exit foils are placed further away, outside of this acceptance range on extensions, thus scattered particles on the windows cannot reach directly the detectors. The telescope arrangement with these long grooves also ensures, that only those scattered particles can reach the detectors, which penetrated perpendicularly the side window. Any secondary scattered particle on the side window or scattered particles from other parts of the setup are shielded.
Comparing this configuration with a scattering experiment on a solid target, the main differences are the energetic width of the beam arriving in the gas volume, and the energy loss of the scattered particles in the side exit foil. As mentioned before in \Sec{sec:gen}, if the reaction cross section and stopping power is nearly constant in the energy range covered by the convolution of the target and beam energy distribution, the determined cross section does not depend on the energetic beam width. However, with a given irradiation energy, this experiments integrate a bigger portion of the excitation function than those on windowless target and sharp beam energy distribution (e.g. \cite{Mohr93-PRC}).
The scattered particles go through the side exit foil towards the detectors, along which they lose energy. The foil thickness gives a lower limit on the scattered particle energy, below which they either stop in the foil, or their energy reduces below the detection threshold. With the 10-$\mu$m thick Al side windows, the scattered alpha particles needs to have at least 3.1 MeV to be detected, however even with this limitation experiments are still feasible, as will be shown in the next section.

Two experiments were performed with the scattering gas-cell introduced in the previous section. In both cases, the $\alpha$ beam was provided by the \mbox{MGC-20} cyclotron of Atomki, and the gas-cell target was installed in the scattering chamber behind an analyzing magnet at beamline 4 \cite{Biri21-EPJP}. This chamber was previously used in number of alpha scattering experiments on solid targets (e.g. \cite{Kiss22-PRC} one of the most recent).
The collimated ion implanted Si detectors of 500 $\mu$m thickness were mounted on a turntable. The energy calibration of these detectors were done using a triple isotope alpha source.
A collimated detector is placed at a fixed angle (30$^\circ$) acting as a monitor detector. The counts in the other detectors were normalized to the monitor, to get their relative yields.
By using this monitor detector approach, neither the absolute charge measurement nor the absolute target density is needed for the data analysis, however, to quote absolute values, the cross section at the monitor angle has to be known.

\begin{figure}[b]
\center
\includegraphics[width=0.95\columnwidth,  angle=0]{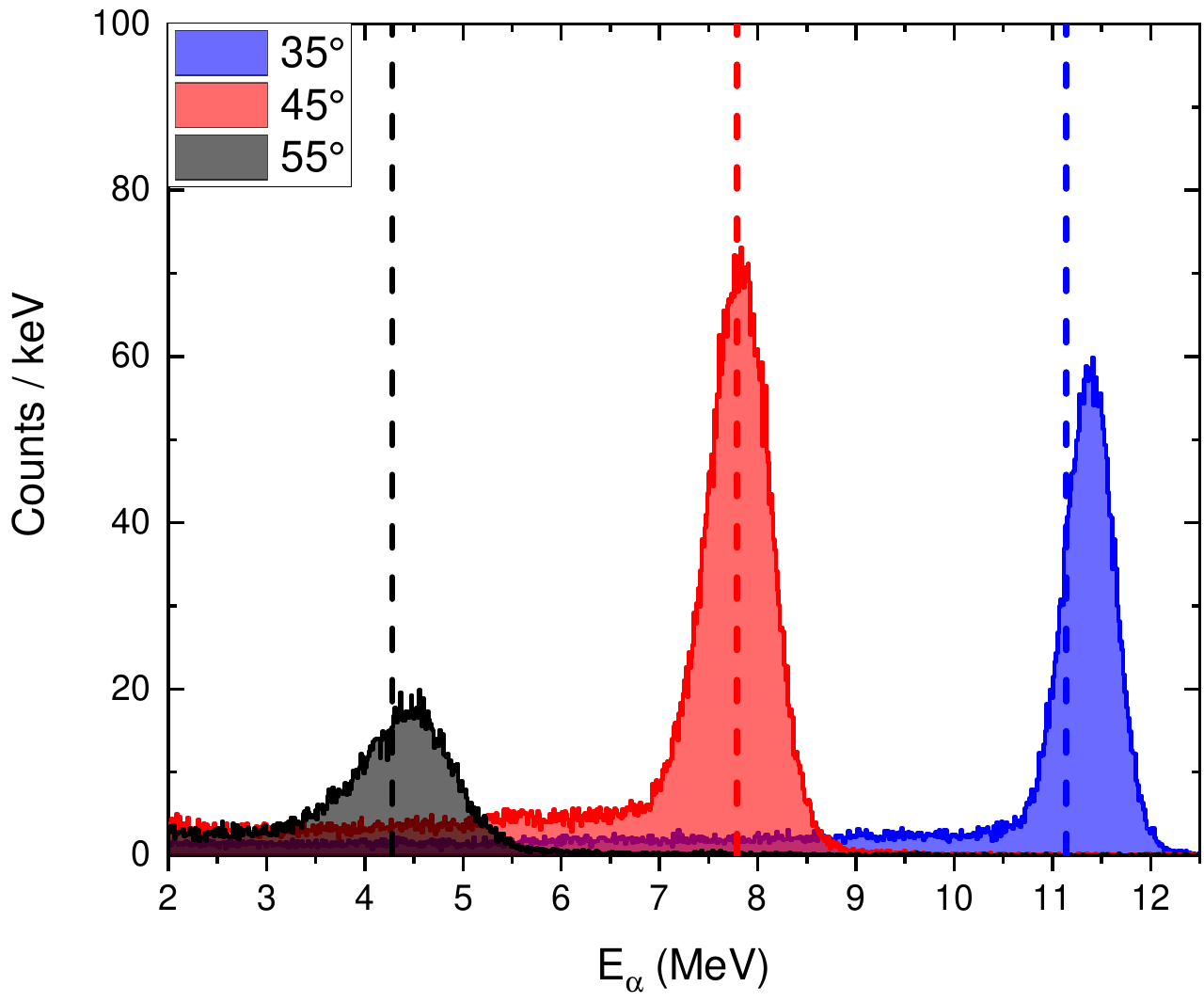}
\caption{\label{fig:he_sc}Scattered alpha energy spectra from the $^{4}$He+$\alpha$ reaction at E$_\alpha =18$~MeV. Three detectors with identical collimators are positioned to the marked nominal laboratory scattering angles. The vertical dashed lines are the calculated energy taking into account the kinematics and the stopping in the exit foil.}
\end{figure}

The pilot experiment was done with E$_\alpha =18.7$~MeV impinging on the scattering gas-cell containing 100~mbar $^{4}$He gas. Inside the gas volume behind the 10 $\mu m$ thick aluminium entrance window, this energy corresponds to  E$_\alpha =18$~MeV. Because the beam and the target are identical, the scattered and recoil particles are indistinguishable.
In addition, because of the reaction kinematics, in the laboratory system only forward scattering is possible. In this case only three detectors were used. The spectra show a distinct peak at each angle, corresponding to elastic scattering (see  \fig{fig:he_sc}). The central energies of the peaks is in fair agreement with the energy expected from the reaction kinematics and the subsequent energy loss of the particles through the side exit foil. The roughly 250 keV mismatch at the lowest angle is attributed to the angular uncertainty of 0.6$^\circ$. By reversing the calculation, the center of the peak gives 34.4$^\circ$ detection angle, within the uncertainty of the system.
At this given beam energy there is a literature dataset \cite{Chien74-PRC} of the reaction.
Using the spectra from this given run the ratio of the peak area can be compared with the literature cross section ratio of $\sigma_{45^\circ} / \sigma_{35^\circ} = 2.05 \pm 0.01$. This can be done, since the collimators can be considered identical, thus the detection efficiency factors out, so do the impinging charge and target density. In the present case, the peak area ratio of $1.73 \pm 0.02$ have to be scaled with $sin(45^\circ)/sin(35^\circ)$ to take into account the conversion between the laboratory and center-of-mass frames, resulting in $2.13 \pm 0.04$. 
The agreement between the ratios shows the usability of the technique. The slight mismatch is due to the above mentioned angular uncertainty and the slight manufacturing difference of the collimators, which causes small efficiency difference.
An experiment with more detectors and better angular calibration can later be used for validation of the setup and data analysis. In addition, it serves as a benchmark towards the study of the $^{3}$He+$^{4}$He scattering.

\begin{figure}[b]
\center
\includegraphics[width=0.95\columnwidth,  angle=0]{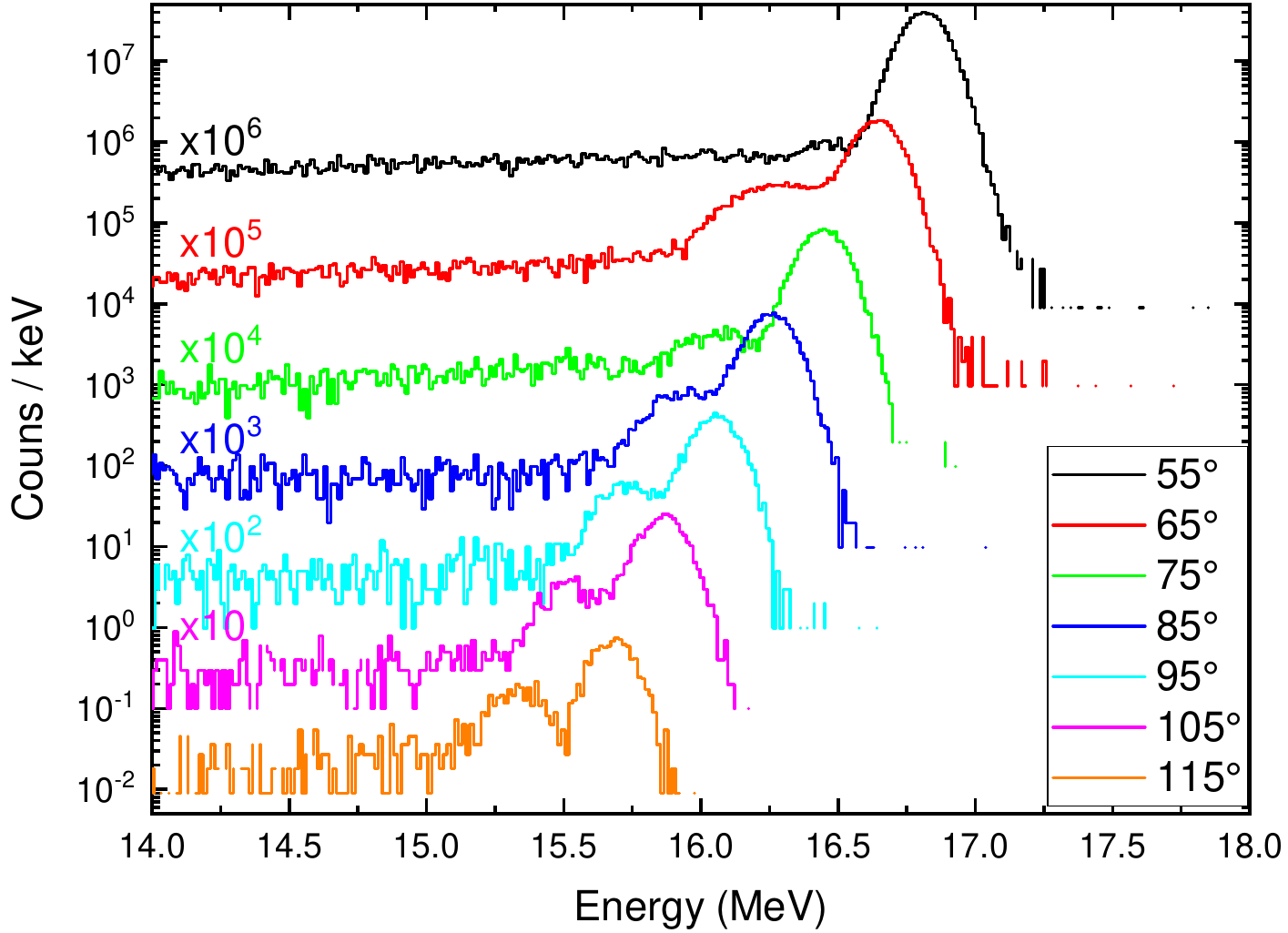}
\caption{\label{fig:Xe_sc}Scattered alpha energy spectra from the $^{124}$Xe+$\alpha$ reaction at E$_\alpha =18$~MeV. The detectors were placed at the marked laboratory scattering angles.}
\end{figure}

The other experiment has been done with the same beam energy and foil windows, however, in this case the gas-cell contained 70~mbar $^{124}$Xe gas. 
A pressure higher than in the activation experiment can be used, since the scattered alpha particles have much higher energy and less energy loss in the gas volume compared to the heavy recoils in the activation experiment.
The alpha scattering spectra were recorded at several angles using all the seven detectors plus the monitor detector shown in \fig{fig:scatt}. In the spectra shown in \fig{fig:Xe_sc}, the signal of the scattered alphas on the heavy gas is clearly visible. The observed peak energy follows the expected behavior by taking into account the energy loss in the entrance foil, the reaction kinematics and the energy loss of the exiting alpha particles. 
The first excited state of $^{124}$Xe is at 352 keV \cite{Katakura08-NDS} which can  be populated in inelastic scattering. This inelastic peak is also visible in the spectra at the kinematically expected energy. Even if the straggling in the exit foil broadens the elastic and inelastic scattering peaks, the separation, thus the peak area determination for cross section derivation is still possible.
For cross section calculation, the normalization to the monitor detector counts, the relative efficiencies of the given detectors and their precise angular calibration have to be implemented, however, it is beyond the scope of the present paper.
Alpha scattering data on noble gas nuclei is scarce in the literature. By such studies the alpha-nucleus optical potential models can be validated, which is a key ingredient for heavy element nucleosynthesis reaction network calculations \cite{Psaltis22-AJ}. 

In summary, the scattering gas-cell performance was tested both on heavy and low mass noble gas target. The recorded spectra show distinct features, which will later be used for scattering cross section determination. It is planned to exchange the side exit window with one with a lower stopping power. Since the side windows are not exposed to the power deposition of the beam, there is a bigger selection of materials to be used. For example silicon-nitrite vacuum windows of 200 nm thickness is routinely used for in-air PIXE measurements \cite{Torok15-NIMB,Aljboor23-JAAS}, serving as exit foils for the low power beam, separating the atmosphere from the beam-line vacuum. Such a foil has more than one order of magnitude lower stopping than that of the presently used 10 $\mu$m Al. This makes it possible to measure at even lower reaction energies and/or bigger angular range, since the scattered alpha particles will lose only 100keV in the exit window.
This is especially important in case of the low mass target, when the scattered alphas have low energy at backward angles.
For example, the lowest center-of-mass energy in case of the $^{3}$He+$^{4}$He scattering, where the scattering cross section still deviates from Rutherford scattering is around 1.2 MeV \cite{Mohr93-PRC}. This correspond to 2.8 MeV $^4$He energy inside the cell filled with $^3$He. To reach this energy after penetrating the 10~$\mu$m Al entrance window, the initial beam energy has to be 4.7~MeV according to SRIM simulations \cite{Ziegler10-NIMB}. These simulations also show, that the energetic width of the beam will be around 35 keV, if started from a monoenergetic beam. Since the maximum energy of the scattered particles are not higher than 2.8 MeV, non of them can penetrate a 10~$\mu$m Al side window. By using the above mentioned 200~nm Si$_3$N$_4$ and considering a minimum 0.4~MeV energy\footnote{This is the lower detection limit set in the present detector setup, however, it can be reduced by further optimization.} of the scattered particles after penetrating the exit window, the scattering cross section can be determined up to 130$^\circ$ c.m. angle.

The lower stopping causes lower straggling, thus the peak widths would also be reduced. This may be advantageous also in case of the heavy target gas for better separation of the inelastic peaks. For those the reduced energy loss is not needed, since even at backward angles the alphas have enough energy to penetrate the aluminium foil. As an example, for the case of $^{124}$Xe, the scattering cross section is dominated by Rutherford scattering below 12 MeV, thus nuclear physics information cannot be deduced from a measurement below this energy. Even at this beam energy, the scattered alpha at 175$^\circ$ will have an energy of about 9.5 MeV after penetrating the exit foil.

\section{\label{sec:sum}Summary and outlook}
In this paper, several stages of the development of thin windowed gas-cell targets were described. Their advantages and disadvantages are shown. Typical experimental and data analysis considerations were detailed, highlighting the valuability and usefulness of such targets. Finally a concrete experimental investigation is shown, which provided new cross section data on the \HeHe reaction in the energy range of the first resonance in the reaction. 

In addition, the development of a gas-cell target for measuring alpha particle scattering cross section on noble gases is presented. It is in developing stage, however, the obtained spectra show a high potential for further use. Especially for alpha scattering on low mass gas, different side window material may be needed for low energy investigations.

Furthermore, based on the experiences gained with the described gas-cells, a gas-cell for in beam $\gamma$-ray spectroscopy is in the design phase. It would be useful, when partial cross sections leading to different excited states in the compound nucleus have to be determined, or \mbox{$\gamma$-ray} angular distribution needs to be measured. In this case a cell with small geometric thickness in the beam direction is desirable, thus the starting point of the $\gamma$ rays can be well defined. The main challenge is the selection of the window material, which is of crucial importance, since parasitic reactions on the window may produce unwanted \mbox{$\gamma$ radiation} spoiling the signal of interest. At the same time, the window have to be thin enough to let the beam penetrate, while standing against the heating effect of the beam power deposition, and the pressure difference.

\section*{Acknowledgments}
We thank the operating crew of the Atomki cyclotron accelerator for their assistance during the irradiations, L\'aszl\'o Palcsu (HUN-REN Atomki) for recording the mass spectra. This work was supported by \mbox{NKFIH} (OTKA \linebreak FK134845, K134197 and K147010), and by the European Union (ChETEC-INFRA, project no. 101008324).

\end{document}